\documentclass[conference,twocolumn,a4paper]{IEEEtran} %a4paper
\IEEEoverridecommandlockouts
\usepackage{setspace}
\usepackage{float}
\usepackage{subfigure}
\usepackage{amsfonts,amssymb}
\usepackage{amsmath}
\usepackage{color}
\usepackage{algorithm}
\usepackage{algorithmic}
\usepackage{geometry}
\usepackage{graphicx}  %插图
\usepackage[mathscr]{eucal}
\usepackage{latexsym}
\usepackage{exscale}
\usepackage{epstopdf}
\def\BibTeX{{\rm B\kern-.05em{\sc i\kern-.025em b}\kern-.08em
		T\kern-.1667em\lower.7ex\hbox{E}\kern-.125emX}}
\usepackage{bm}
\usepackage{cases}
\usepackage{epsfig}
\usepackage[center,small]{caption}
\usepackage[verbose,nospace,sort]{cite}
\usepackage{tabularx}
\usepackage{multirow}
\usepackage{latexsym}
\usepackage{balance}
\usepackage{url}
\usepackage{mathtools}
\usepackage{tablefootnote}
\usepackage{hyperref}
\usepackage{graphicx}
\usepackage{subcaption}
\usepackage{cleveref}
\title{An Overview of Cellular ISAC for Low-Altitude UAV: New Opportunities and Challenges }

\author{\IEEEauthorblockN{Yuxuan~Song, Yong~Zeng,~\IEEEmembership{Fellow,~IEEE,} Yuhang~Yang, Zixiang~Ren, Gaoyuan~Cheng, \\Xiaoli Xu,~\IEEEmembership{Member,~IEEE,} Jie~Xu,~\IEEEmembership{Fellow,~IEEE,} Shi Jin,~\IEEEmembership{Fellow,~IEEE,} and Rui Zhang,~\IEEEmembership{Fellow,~IEEE} }
\thanks{
Yuxuan Song, Yong Zeng (corresponding author), Yuhang Yang, Xiaoli Xu, and Shi Jin are with Southeast University, China; Yong Zeng is also with the Purple Mountain Laboratories, China.

Zixiang Ren, Gaoyuan Cheng and Jie Xu are with The Chinese
University of Hong Kong, Shenzhen, China.

Rui Zhang is with The Chinese University of Hong Kong, Shenzhen, Shenzhen Research Institute of Big Data, and National University of Singapore.
}

}

\begin{document}

\maketitle

\begin{abstract}

Low-altitude unmanned aerial vehicles (UAVs) are expected to play an important role in future wireless networks, either as aerial base stations (BSs) or aerial users connected to the cellular network. In addition, integrated sensing and communication (ISAC) has been identified as one of the six usage scenarios for the forthcoming sixth-generation (6G) mobile networks, aimed at improving network functionalities and realizing situational awareness of the physical world. While most existing research efforts focus on terrestrial two-dimensional (2D) communication and sensing, UAV as an aerial platform offers a new degree of freedom for designing three-dimensional (3D) air-ground (AG) ISAC networks. In this article, we provide an overview of cellular-connected UAV ISAC, by elaborating the UAV's roles as a target to be sensed and as an aerial anchor to provide sensing functionality, respectively. In particular, we pay attention to the network coverage issue and topics specific to UAV networking, emphasizing the new opportunities as well as unique challenges to be addressed.

\end{abstract}


\pagestyle{plain}

\section{Introduction}
Low-altitude unmanned aerial vehicles (UAVs) typically refer to those of small and medium size operating at no higher than 3 kilometers (km) above the ground. They are crucial to the expansion of human activities from ground to low-altitude airspace such as the development of low-altitude economy. Among the various available wireless technologies for supporting low-altitude UAV operations, the cellular-connected approach is considered to be the most effective to realize large-scale UAV deployment and enable beyond visual line-of-sight (BVLoS) UAV operation with essentially unlimited communication range \cite{zeng2019accessing}. In particular, by exploiting the worldwide accessibility and advanced transmission technology of cellular networks, efficacy of on-demand UAV services such as traffic control, aerial imaging and UAV-enabled airborne communication is anticipated to be amplified to the maximum, especially in terms of operation range and data rate. The technique specifications for 5G new radio (NR) support for UAVs have been released by 3GPP in \cite{3gpp18}. At the same time, integrated sensing and communication (ISAC) has been viewed as a powerful enabler for future 6G networks, which is aimed at improving network functionalities and enhancing our perception of the physical world\ \cite{ITU2030}. When we take sensing-related demands and corresponding metrics into consideration, the paradigm of cellular-connected UAVs shows even more dominating superiority.

The interpretation for the above statement is twofold. First, there have been growing interests/needs for UAV detection, localization, classification and tracking. On the one hand, non-cooperative UAVs tend to pose huge threat to public safety and reliable detection of them is the prerequisite for countermeasures. On the other hand, communication services for aerial user equipments (UEs) can be greatly enhanced by sensing capability. For example, UAV location and trajectory-related information could facilitate beam alignment process or enable the acquisition of more accurate channel state information (CSI)\cite{zeng2023ckm}. Inter-connected cellular networks deployed worldwide are perfectly suitable for the aforementioned tasks with inherent data sharing and fusion capabilities. Second, with continuous cost reduction and device miniaturization, UAVs are endowed with the capability to provide more sensing-related services including ground vehicle navigation, synthetic aperture radar (SAR) imaging and so on. By integrating communication and sensing hardware into cooperative UAVs, an aerial ISAC network capable of both internal (UAV-to-UAV) and external (UAV-to-BS) data exchange can be established, expanding our vision from two-dimensional (2D) to three-dimensional (3D) where the desired line-of-sight (LoS) channel characteristic could be fully exploited.

Based upon the above reasoning, we regard cellular networks as a key enabler for enhancing the efficacy of UAV ISAC applications. In other words, only through network-level connection can the full potential of UAV ISAC be completely unleashed. As illustrated in Fig. 1, from the wireless sensing perspective, UAVs may play different roles in the network, i.e., as aerial targets to be sensed or as aerial anchors to provide ISAC capabilities from the sky, which brings both new challenges and opportunities. Compared with conventional UAV communications or ground ISAC systems, UAV ISAC systems possess several unique characteristics:

\begin{figure*}
	\centering
	\includegraphics[scale=0.45]{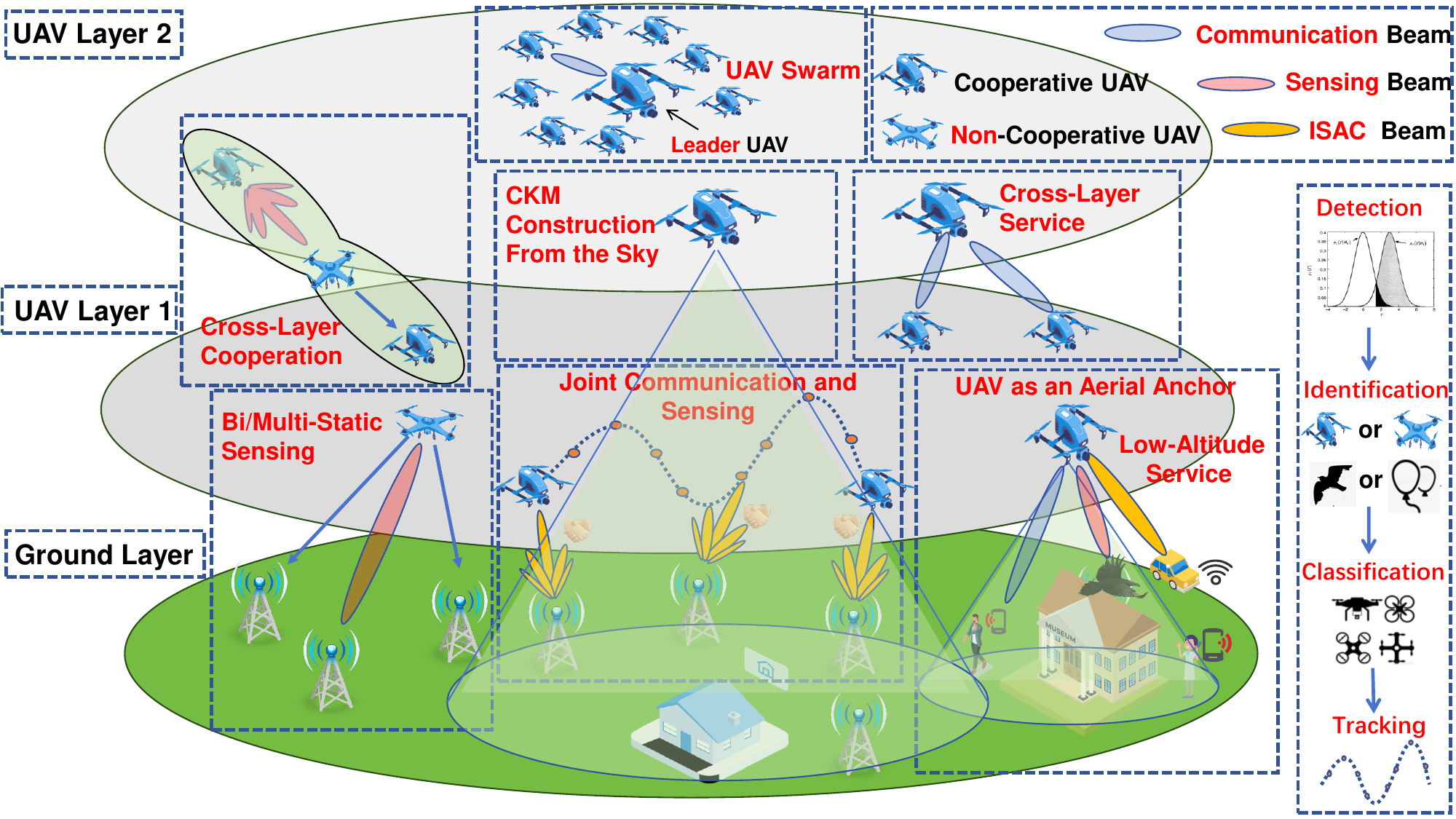}
	\caption{An illustration of cellular-connected UAV ISAC.}
	\label{F:NewChara}
\end{figure*}

\begin{itemize}
	
\item[(1)] \textit{Complication for Coverage:} Different from conventional UAV communication which only considers communication coverage by virtue of metrics such as signal-to-interference-plus-noise ratio (SINR), UAV ISAC should also concern sensing coverage with sensing metrics like signal-to-clutter ratio (SCR). Meanwhile, as opposed to terrestrial ISAC network that only needs to consider coverage on 2D planes, the additional spatial dimension introduced by UAV's mobility transforms the concept of coverage from area to volume and correspondingly calls for 3D beamforming, which tends to further twist the shape of both communication and sensing coverage. 

\item[(2)] \textit{Multi-Layer Architecture:} Geographically separated BSs at the terrestrial layer can act as a network for sensing non-cooperative UAVs and providing ISAC service for cooperative ones. Meanwhile, cooperative UAVs can be deployed at different layers to take on missions with different optimum operation altitudes to optimize the ISAC service quality from the sky.

\item[(3)] \textit{Low Detectability:} As opposed to targets in ground ISAC networks with large radar cross section (RCS) such as cars or humans, consumer-grade UAVs are classified as typical low-altitude, low-speed and small-RCS targets due to their limited size and mostly non-metallic structural components, leading to their low detectability and trackability in general. Consequently, multi-pulse accumulation technology is made indispensable for improving the SINR and SCR of the reflected echo.

\item[(4)] \textit{Airframe Influence:} The shape of the UAV also plays a significant role, affecting multiple aspects including RCS signature, airframe shadowing\cite{AGChannel} and flying energy consumption. In particular, airframe shadowing is an impairment unique to air-ground (AG) channel and has a huge adverse effect on the communication and sensing quality. 
%As the commercial UAV market further progresses, more diverse models of UAV are expected to appear, which demands for a deeper understanding of their respective characteristics.

\end{itemize}

In this article, we provide an overview of UAV ISAC networks by holding on to the cellular-connected paradigm and regard it as a foothold, emphasizing the significance of sensing for UAVs and their role as aerial anchors. Meanwhile, we pay special attention to the network coverage issue and topics specific to networking.

\section{UAVs as Sensing Targets}

One characteristic feature posed by the introduction of UAV to cellular network is that the sensing for UAV, whether cooperative or not, is of paramount importance since it is the precondition for both countermeasures against non-cooperative UAVs and the successful cooperation among cooperative ones. In this section, we discuss three concepts specific to this topic:
\begin{itemize}
	\item[(1)] \textit{Detection and Estimation:} “Detection” means reporting the presence of targets in the airspace of interest, usually realized with constant false alarm rate (CFAR) algorithm together with preliminary parameter estimation including range, angle of arrival (AOA), and radial velocity. 
	
	\item[(2)] \textit{Identification and Classification:} “Identification” refers to distinguishing between UAVs and other aerial targets such as birds or insects, followed by classifying the detected UAV among multiple types such as fixed-wing, quadcopter or octocopter, and even determining its exact model. 
	
	\item[(3)] \textit{Tracking:} The last procedure “Tracking" stands for timely updating the UAVs’ trajectory and state information without ambiguity, in the sense that the tracker must be capable of differentiating flight paths belonging to different UAVs with developed authentication mechanism. 
\end{itemize} 
%In this section, we discuss these topics in detail.

\subsection{RCS and Detection}
%RCS-based detection
RCS is one of the most important characteristics of a target in the context of sensing. According to detection theory, the shape of the probability distribution function (PDF) of the target’s RCS directly affects the structure of the detector. The more accurate the PDF is fitted, the better performance the system might achieve. Detailed description and characteristic analysis of the target RCS requires integration of the data obtained by various means including theoretical calculation, anechoic chamber measurement and outfield measurement.
As the size of the UAV grows and the signal bandwidth of the cellular system increases, a single UAV might take up more than one resolution cell, turning from a point target to an extended target. This brings new modelling issues as well as opportunities in terms of RCS analysis.

In addition, target’s RCS in bi-static sensing architecture tends to be more complex than that in mono-static case, since it depends on the bi-static angle, defined as the angle between the transmitter-to-target LoS link and target-to-receiver LoS link. Consequently, further research is needed for capturing UAV’s bi-static RCS characteristic, especially the relationship with its mono-static counterpart \cite{willis2005bistatic}.

\subsection{Identification Methods}
An important task for effective radar-based identification of UAVs, is the discrimination between UAVs and birds or insects. This is a non-trivial task since they may share comparable RCS and nearly identical motion patterns in both speed and trajectory, which might lead to significant false alarm rates. Fortunately, their respective Micro-Doppler signatures are fairly distinct, rendering reliable classification possible.

As long as the sensing system is sensitive enough, in the sense that the frequency resolution is sufficiently high to capture the Micro-Doppler signatures that reveal features such as blade flashes and propeller rotation, it is expected to be capable of discriminating UAVs from other aerial targets whose micro-motion frequency are much lower. However, most existing literature simulates the received echo with radar-specific waveforms. It remains unanswered whether the same performance can be achieved with ISAC waveform like orthogonal frequency division multiplexing (OFDM).

In addition, by utilizing the large array aperture and advanced hardware of modern cellular BSs, more complicated yet more accurate identification methods could be applied like inverse synthetic aperture radar imaging and polarization.

\subsection{Joint Communication and Tracking}

The basic philosophy of tracking is to associate the prediction value based on previous measurements with new measurements so that each tracked target’s trajectory information can be correctly updated. To be specific, targets are illuminated by sensing beams periodically in order that the discrete samples (observation values) of their continuous trajectory are recorded. The tracker then assigns these detections to existing tracked targets or launches a new tracking mission based on data association algorithms like global nearest neighbor (GNN). The real difficulty of long-term, steady-state multiple aerial target tracking lies in the successful discrimination and trajectory association among multiple UAVs, especially when UAV swarms are deployed. Compared to detection and identification, tracking is a far more complex and dynamic topic that is much more sensitive to clutter interference, which can be easily generated by high-rise buildings or trees in urban scenarios. 

Last but not least, the high mobility of UAVs puts more stringent demand on the refreshing rate of state information, which is constrained by various factors including scanning period, computing speed and the transition mechanism between searching and tracking. All these topics require further research and standardization efforts.

\begin{figure}[H]
	\centering  %图片全局居中
	\subfigbottomskip=2pt %两行子图之间的行间距
	\subfigcapskip=-5pt %设置子图与子标题之间的距离
	\subfigure[Successful Tracking]{
  
      \includegraphics[width=8cm,height=7cm]{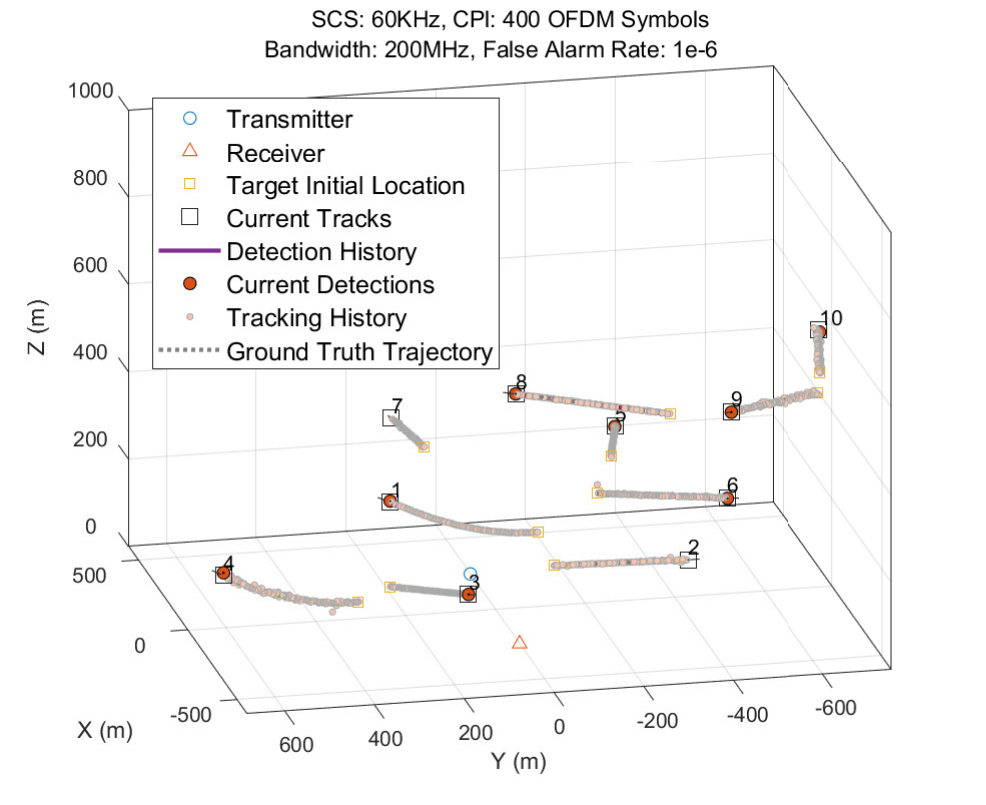}}
        
	\subfigure[Trajectory Ambiguity]{
		\includegraphics[width=8cm,height=7cm]{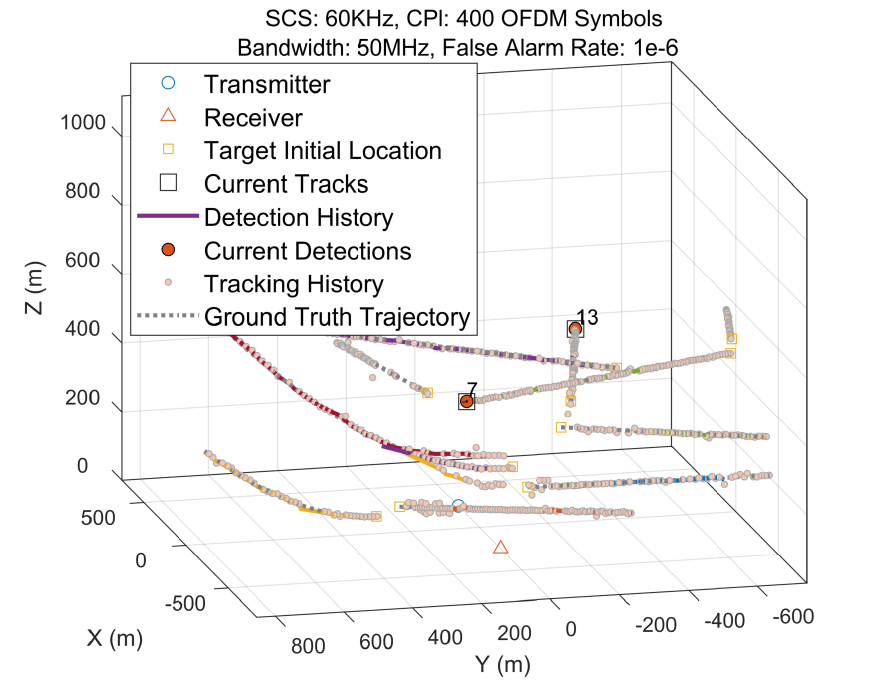}}
%	  //
%	\subfigure[fig3]{
%		\includegraphics[width=0.48\linewidth]{Pictures2/Tracker.png}}
%	%\quad
%   //
%	\subfigure[fig4]{
%		\includegraphics[width=0.48\linewidth]{Pictures2/Tracker.png}}
	\caption{Tracking Scenario Visualization.}
\end{figure}

\subsection{Simulation Results}

Inspired by the mobility model considered in 3GPP simulations and the superiority of bi-static sensing architecture, we consider an ISAC scenario with two BSs acting as a bi-static cooperation pair. The transmitter BS sends communication signal for 10 randomly generated UAVs while the receiver detects and tracks them by processing their overlapped echoes. Target parameters including flying altitude, speed and moving patterns (constant velocity and constant acceleration) are derived from\cite{3GPP}. During each coherent processing interval (CPI), the aggregated echoes are first processed in the range-doppler domain and then to a CFAR detector where the existence of these UAVs is reported. The AOAs are calculated with multiple signal classification (MUSIC) algorithm and the UAVs' 3D locations are subsequently determined. These locations are further fed into a tracker based on kalman filtering and GNN algorithm.

Fig. 2(a) is an overall visualization of the simulation scenario where SCS stands for subcarrier space. It is observed that all the UAVs can be reliably detected based on prescribed false alarm rate and the tracker successfully distinguished the UAVs' trajectories without ambiguity. When the bandwidth is too small, the data association algorithm fails to function and trajectory ambiguity occurs, as shown in Fig. 2(b). Meanwhile, it could be seen from Fig. 3 that the allocated bandwidth has a major impact on tracking accuracy while the variation in false alarm rate has a comparatively milder effect.

\begin{figure}[H]
	\centering
	\includegraphics[width=8cm,height=6.7cm]{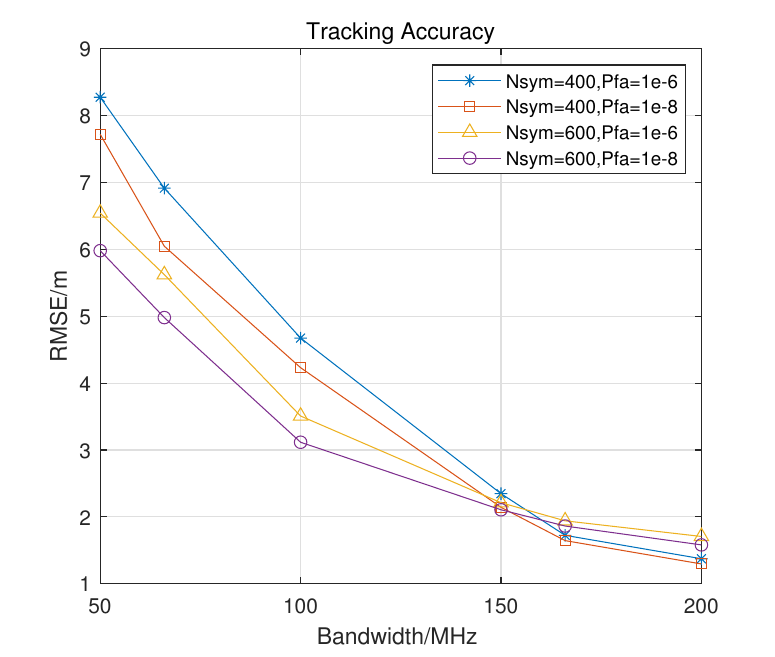}
	\caption{UAV tracking accuracy for different bandwidth and false alarm rate. }
	\label{F:ModelofRCS}
\end{figure}

\section{UAVs as Aerial Anchors}

In this section, we switch our perspective to treating UAVs as aerial anchors providing sensing functionalities from the sky.

\subsection{Classification for Cooperative UAVs}
Being cooperative can significantly improve the sensing capabilities of UAV anchors. Cooperative UAVs can be loosely classified into three types, namely aerial BS, aerial UE and aerial ISAC platform. Aerial BSs focus on providing communication-related services such as mobile relay and hotspot offloading for both terrestrial and aerial UEs. Aerial UEs on the other hand, perform certain particular tasks like video streaming, goods delivery and SAR imaging while maintaining communication with terrestrial or aerial BSs from which the control and non-payload communication (CNPC) signals are sent \cite{zeng2019accessing}. Lastly, aerial ISAC platforms equipped with ISAC transceivers are capable of providing both communication and sensing services, which demands for more advanced beam management algorithms and higher mobility given the complication in service requirements and more stringent connectivity constraints. All UAVs should be properly sensed regardless of their types in order to ascertain the safety of the network. Meanwhile, all cooperative UAVs, which are the focus of this section, should be connected to terrestrial or aerial BSs so that the controllability requirement is satisfied.

\subsection{UAV as an ISAC Platform}

Before discussing the roles and potentials of aerial ISAC platforms in detail, one important question that needs to clarify is what makes it necessary or desirable to equip a cooperative UAV with ISAC devices. In other words, what's the motivation for integrating ISAC capability into UAV instead of other types of vehicle. In order to address this question, it is imperative to emphasize the favorable characteristics of UAV as an aerial platform:

\textit{LoS-Dominated Channel}: As opposed to the case in modern communication where the multi-path components act as an extra exploitable degree of freedom for system performance enhancement, echoes from “non-target" scatterers, also known as clutters, are always detrimental to sensing. The fewer the NLoS components, the more favorable it is for sensing. Consequently, the LoS-dominated characteristic of AG channels makes UAV inherently suitable for ISAC services.
	
\textit{Flexible Mobility}: As pointed out in various previous articles\cite{lyu2022joint}, the mobility of the UAV acts as the core advantage of UAV-enabled use cases. A concept tightly associated with this property is trajectory optimization, which aims to fully exploit this extra degree of freedom for performance enhancement.
	
\textit{Re-Organization Capability}: This feature relies heavily on the mobility property yet is also a fundamental advantage compared to static ground BSs. The aerial ISAC network can swiftly reconfigure its topology according to on-demand service requirements, thus improving both the service's efficacy and efficiency in general.

Other types of cooperative UAVs like aerial BSs might also possess elementary sensing capability in order to meet the basic requirements on obstacle avoidance. However, this is usually achieved by adding extra radar transceivers. Communication and sensing functionality on an aerial ISAC platform should share the common time, frequency and spatial resources based on a set of unified ISAC transceiver. One should pay special attention to the deployment scenario to appropriately determine the major task (communication or sensing) of a particular aerial ISAC platform since it directly affects the resource allocation on-board. In the meantime, it should be pointed out that there are several challenges for using UAVs as aerial ISAC platforms. Here we exemplify with three issues:

\textit{Connectivity Constraint}: For terrestrial BSs, their mutual communications mainly rely on wired links like optical fiber, which is apparently not feasible for an aerial BS or aerial ISAC platform. Therefore, an aerial ISAC platform itself should be cellular-connected so as to qualify as an anchor, which brings additional connectivity constraints for optimizing their trajectories and subsequently influence coverage area. 
	
\textit{Computation Offloading}: When being assigned computation-intensive tasks like pattern recognition or CFAR detection, it is evidently an unreasonable decision to have all the computation done on-board considering the limited energy and computing resources. Computation offloading is considered an effective approach to solve this issue \cite{meng2023uav}. By offloading the raw or pre-processed data to the associated BSs or edge servers where the computation is executed in parallel, the resource overhead on-board can be significantly reduced.

\textit{Clutter Mitigation}: Ground clutter interference is an unique and inevitable impairment to AG sensing link and demands for cancellation techniques like moving target indication (MTI).

\subsection{Association with CKM}
Channel knowledge map (CKM)\cite{zeng2024tutorial} is a site-specific database that learns the physical channel parameters such as path loss, delay and AoA. It acts as a digital twin capable of providing a priori channel information for both communication and sensing services. From the viewpoint of engineering, the motivation for the construction of CKM is to record the reusable channel information that is previously squandered and ultimately replace stochastic-based analysis with deterministic approaches. Mathematically speaking, the fundamentally rigorous idea behind CKM is to map the 3D spatial locations or virtual locations into vectors belonging to a high-dimensional vector space. Thus, CKM possesses high research significance both in theory and in practice.

There are two major tasks in CKM, namely CKM construction and CKM application. For construction, the more accurate the receiver is localized, the higher resolution the database can acquire. High localization precision is comparatively easier for aerial receivers due to less clutter interference and the fact that cooperative UAVs have steadier moving patterns and more predictable trajectory than ground UEs. What’s more, thanks to the LoS-dominated characteristic of AG propagation channel, both the data volume and the measurement complexity of aerial CKM are significantly smaller than that of their terrestrial counterparts. In terms of application, CKM can help judge the existence of LoS links and subsequently assist in determining the optimum beamforming vectors. It can also facilitate the implementation of clutter mitigation as previously specified. All these mentioned above highlight CKM's inherent compatibility with UAV-enabled use cases.

\section{Networking Perspective}
With cellular-connection as the foundation of various UAV-enabled ISAC use cases, in this section, we discuss several topics specific to networking.

\subsection{Multi-Static Sensing Architecture}
A prominent advantage of using cellular network for aerial target sensing (e.g., target detection, parameter estimation, long-term tracking and imaging) over conventional mono-static radar systems is network-level cooperation. Geographically separated ground BSs can form a multi-sensor network capable of flexibly adjusting the transmitter-receiver pairs so that seamless aerial sensing coverage can be achieved. In addition, neighboring BSs working in cooperative mode can neatly regulate both sensing type (active/passive) and sensing mode (mono-static/bi-static) so as to cater to different sensing needs. For localizing cooperative UAVs or aerial UEs sending CNPC or payload communication signal, terrestrial BSs can work in a passive manner where the received signals’ time difference of arrival (TDOA) or AoA could be exploited. For non-cooperative UAVs, on the other hand, ground BSs can actively radiate ISAC signal and jointly process the echo to acquire the exact number and parameters of the targets. Moreover, by fusing multiple measurements from different sensors, system performance regarding localization accuracy and full observability of the target state is expected to be substantially enhanced.

One major challenge to ensure accurate localization lies in the non-linear nature of solving ellipsoid equation, which can be both computationally-demanding and error-prone. Meanwhile, there exists the issue of ghost intersections in multi-sensor-based localization approaches, which can lead to either false alarms or missed detections.

\subsection{Multi-Layer Architecture}
Intuitively speaking, deploying UAVs with distinct operation tasks or mobility attributes at different altitudes can consolidate the order of the network and generally diminish the risk of collision, subsequently simplifying the procedure for control and monitoring. In the context of ISAC, multi-layer architecture can be beneficial to both communication and sensing and offers more degrees of freedom for network design.

For communication, aerial BSs can be deployed at higher layers and provide communication services for both terrestrial and aerial UEs to tackle with potential coverage holes by ground BSs. Besides, the additional computing resources available for processing offloaded data can assist in enhancing the aerial network's capacity in terms of the number of UAVs and communication throughput. For sensing, on the other hand, the optimum operation altitude for diverse sensing tasks could be different and multi-layer deployment can prevent potential performance degradation. Furthermore, UAVs operating at adjacent layers can cooperate in a cross-layer manner to form bi-/multi-static radars where the selection of cooperation pairs tends to be highly dynamic and flexible, which is particularly beneficial to sensing aerial non-cooperative objects.

\subsection{UAV Swarm}

Due to the inherent size, weight, and power (SWAP) constraints, a single UAV has rather limited serving coverage and endurance. UAV swarm, on the other hand, coordinate multiple UAVs for a common task by forming a self-organizing network. In the context of ISAC, sensing data of each wingman UAV in the swarm can be transmitted to and jointly processed at the leader UAV or ground BSs, obtaining information-level or even signal-level fusion gain\cite{meng2023uav}. Meanwhile, thanks to its flexible self-organizing capability, the network topology can be swiftly reconfigured according to environmental information and adjustments in sensing and communication requirements. Lastly, in order to combat the constraint of limited on-board energy, a UAV swarm can dynamically alter its accessed members without sacrificing the continuity of ISAC services, achieving better robustness and endurance in general.

From the viewpoint of communication, it would be inefficient and wasteful of system resources to connect each UAV in the swarm directly with ground BSs due to their large number. On the sensing side, however, terrestrial BSs ought to have the ability to characterize the 3D position and trajectory of each UAV individually, despite their close separation. For the aim of detection, this necessitates the use of super-resolution signal processing algorithms such as MUSIC. In terms of tracking, conventional GNN algorithm suitable for tracking sparsely distributed targets might no longer be reliable in such a situation and more complicated data association algorithms such as joint probabilistic data association (JPDA) should be applied so as to prevent trajectory ambiguity.

\begin{figure}[H]
	\centering  %图片全局居中
	\subfigbottomskip=2pt %两行子图之间的行间距
	\subfigcapskip=-5pt %设置子图与子标题之间的距离

	\subfigure[GDOP with TDOA as the dominant measurement error]{
		\includegraphics[width=8cm,height=6cm]{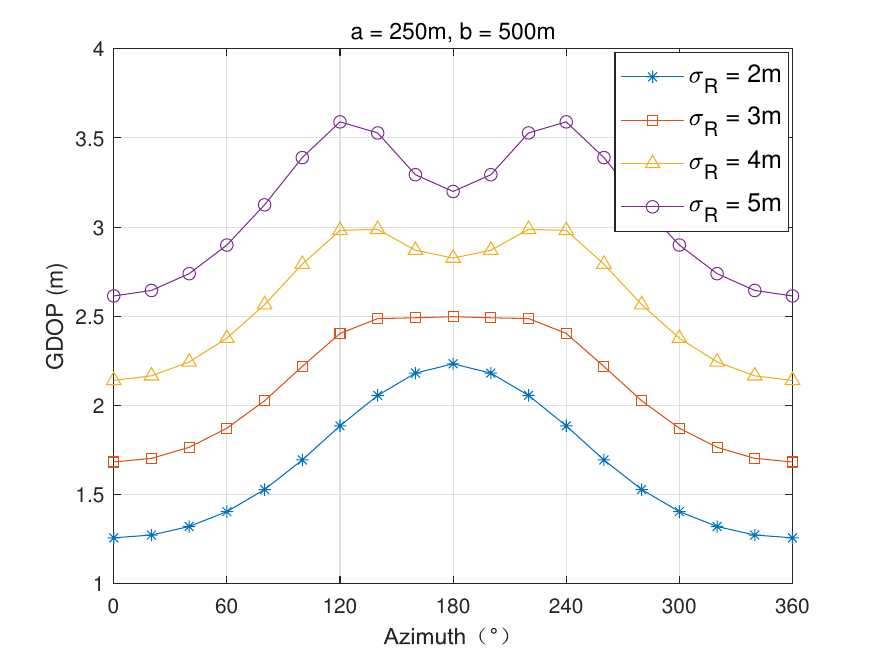}}
    
	\subfigure[GDOP with AOAs as the dominant measurement error]{
		\includegraphics[width=8cm,height=6cm]{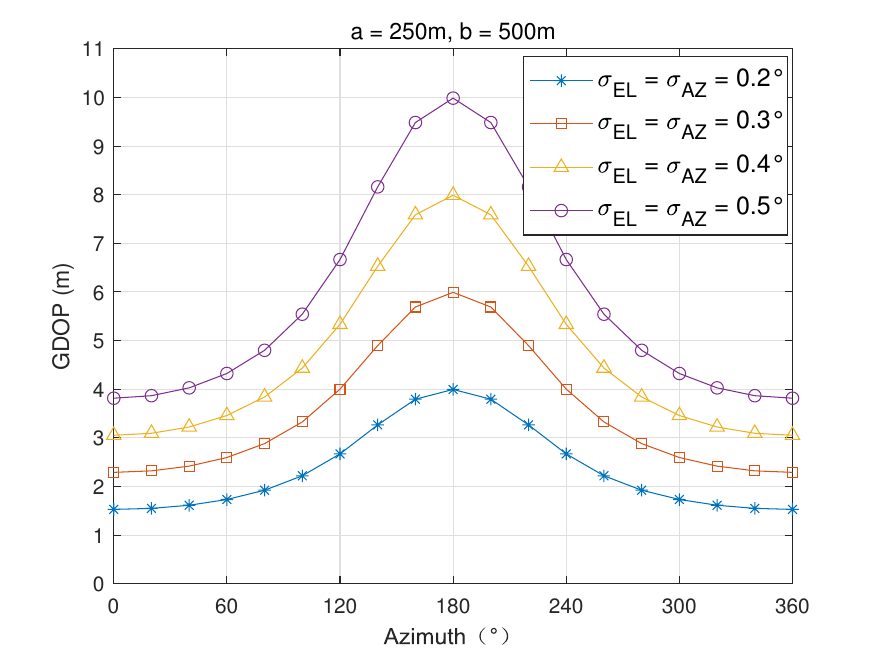}}  

%         \subfigure[fig1]{
%		\includegraphics[width=4.5cm,height=4cm]{Figure/A0.707_Final.eps}}   
	
%         \subfigure[fig2]{
%		\includegraphics[width=4.5cm,height=4cm]{Figure/T0.707_Final.eps}}
%	  //
%	\subfigure[fig3]{
%		\includegraphics[width=0.48\linewidth]{Pictures2/Tracker.png}}
%	%\quad
%   //
%	\subfigure[fig4]{
%		\includegraphics[width=0.48\linewidth]{Pictures2/Tracker.png}}
	\caption{GDOP with different measurement errors.}
\end{figure}

\subsection{Geometry-Related Issue}
The cooperation among neighboring BSs brings extra issues that don’t exist in mono-static sensing mode. For example, in order to successfully detect and localize multiple aerial targets, cooperative BSs should be aware of the LoS distance to one another. Additionally, the large-scale coordination among networked ISAC transceivers requires accurate time and phase synchronization to avoid distortion. Once the UAV exceeds the ISAC coverage region of currently associated BS, there must be a switching mechanism to set up new cooperative relationships so as to provide seamless communication and sensing services. On the other hand, complication in geometry also introduces new designing perspectives and optimization objectives. A featured concept in bi-/multi-static sensing, whose detailed architecture is shown in Fig. 5, is geometric dilution precision (GDOP)\cite{kanhere2021target}, which expresses the sensitivity of target localization accuracy to measurement error. GDOP can assist with selecting the optimum transmitter/receiver location, which is of great utility when optimizing sensing architecture. Fig. 4(a) and 4(b) respectively show the curves of GDOP variation over azimuth angle with measurement errors mainly in TDOA and AOAs. It can be concluded that the closer the target is to the receiver, the less susceptible the localization accuracy is to AOA measurement errors. which is intuitively satisfactory with the azimuth angle equaling to 180 degrees as the worst case. While the case in TDOA is more complicated where the GDOP might change non-monotonically as the azimuth angle increases from 0 to 180 degrees.

\begin{figure}[H]
	\centering
	\includegraphics[width=9cm,height=5cm]{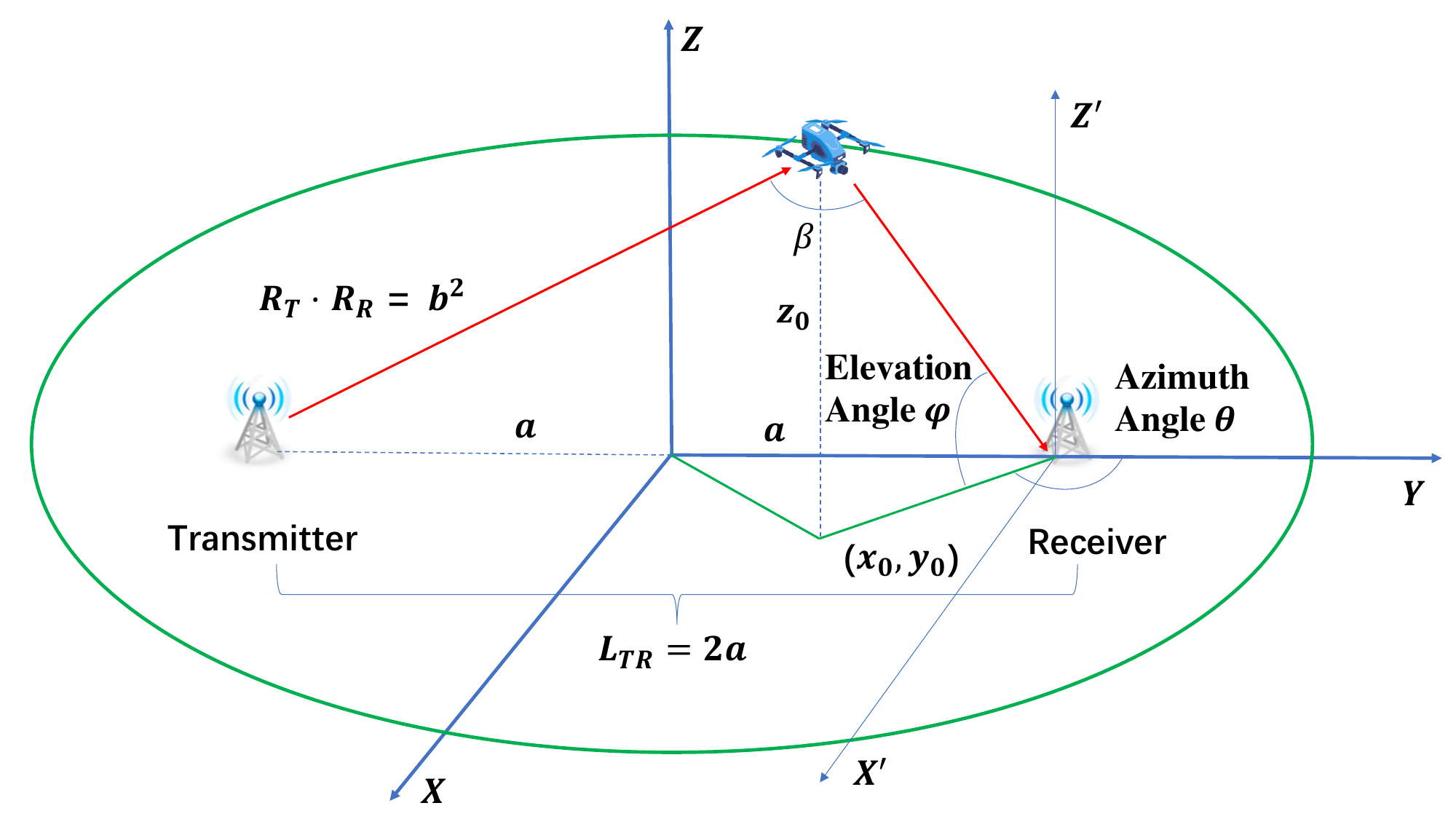}
	\caption{Geometry of Bi-static sensing for UAV.}
	\label{F:ModelofRCS}
\end{figure}

\section{Coverage analysis}
Coverage is undoubtedly the most important issue when building a UAV ISAC network\cite{li2023towards}. Owing to the intrinsic difference between sensing and communication in terms of coverage requirement, it is of paramount importance to capture their respective major characteristics. For example, unlike the procedure of initial communication access which is more of a two-way handover process, sensing tends to be one-way and emitter-centered. In addition, sensing is discrete and periodic in nature while communication must be “always-on”. This fundamental difference could pose difficulties in designing fully-unified ISAC waveform and the beam management\cite{golzadeh2023downlink} issue coupled with it.

\subsection{Ground BSs for Aerial Coverage}

Ground BSs are static in nature and therefore the degrees of freedom for coverage optimization lie in their placement and beam management mechanisms. The interpretation for BS placement optimization is to decide the optimum deployment topology of cooperative BSs regarding distances, relative angles, density and the cooperation mechanism among them. This task focuses on network-level analysis and correspondingly requires probabilistic analytical tools such as GDOP discussed in the earlier section. Beam management on the other hand, pays attention to waveform design and resource allocation, with metrics like SINR as objective function to be maximized. As discussed in Section IV, one favorable advantage of using cellular network for UAV sensing is the flexibility of bi-/multi-static architecture. Based on bi-static radar equation, the SNR equipotential surface is a standard Oval of Cassini when the transmitter radiates sensing signal isotropically\cite{willis2005bistatic}. When taking the BS’s 3D beampattern into consideration, the radiation power becomes a function of azimuth and elevation angles. Consequently, the actual coverage turns into a deformed Oval of Cassini, illustrated in Fig. 6 with both transmitter-receiver distance and geometric average distance equaling to 500 meters. Compared to 2D beamforming, 3D beamforming tends to further twist the shape of coverage area and cause the objective function to be more mathematically irregular, posing difficulties when devising optimum beamforming algorithms. Meanwhile, it must be pointed out that severer inter-cell and intra-cell interference are bound to appear as more beams are steered towards the sky, making advanced multiple access and interference cancellation techniques indispensable. 

\begin{figure}[H]
	\centering  %图片全局居中
	\subfigbottomskip=2pt %两行子图之间的行间距
	\subfigcapskip=-5pt %设置子图与子标题之间的距离
	\subfigure[Standard 2D Oval of Cassini]{
		\includegraphics[width=4.3cm,height=4cm]{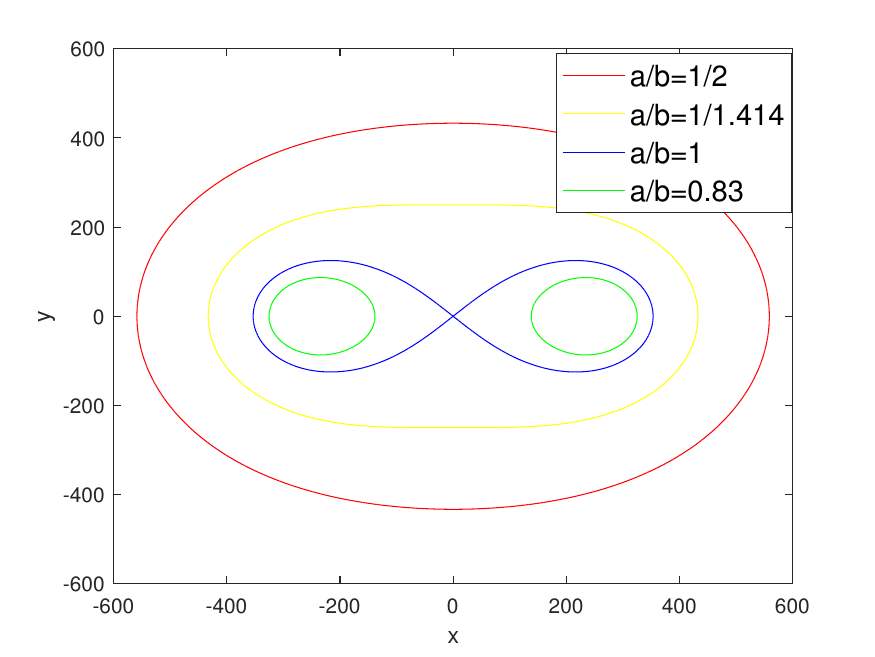}}     
	\subfigure[Standard 3D Oval of Cassini]{
		\includegraphics[width=4.3cm,height=4cm]{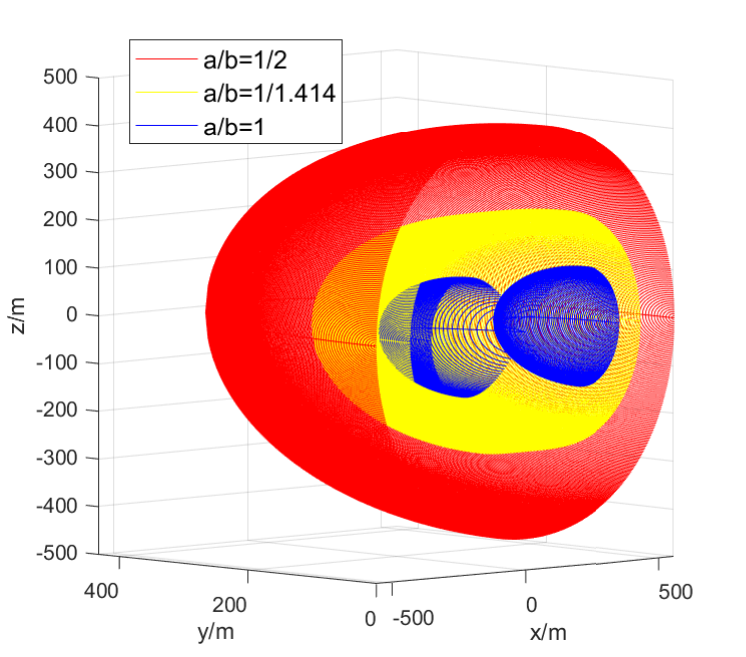}}
        \subfigure[Deformed Cassini based on BS Beampattern (1)]{
		\includegraphics[width=4cm,height=3cm]{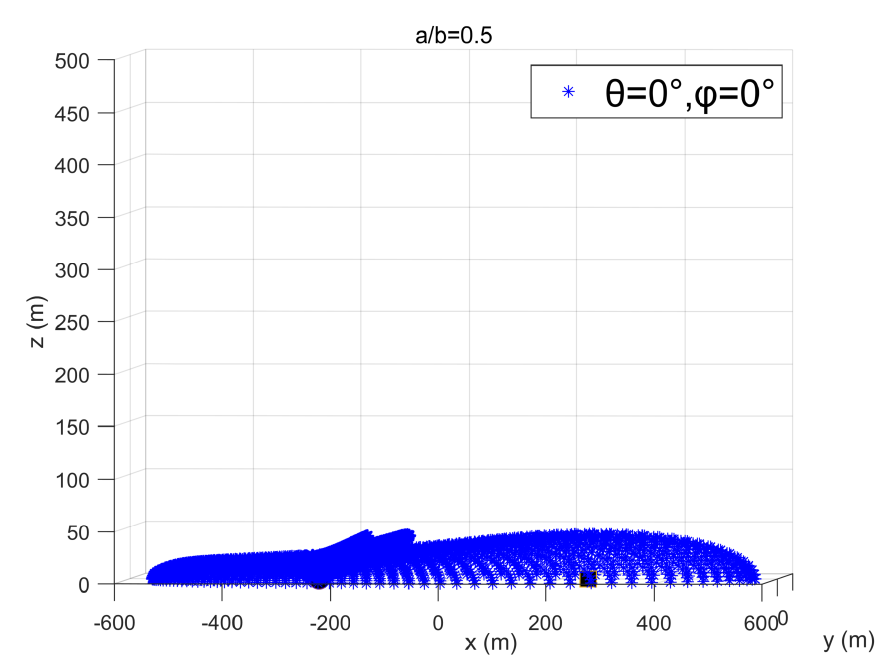}}
        \subfigure[Deformed Cassini based on BS Beampattern (2)]{
		\includegraphics[width=4cm,height=3cm]{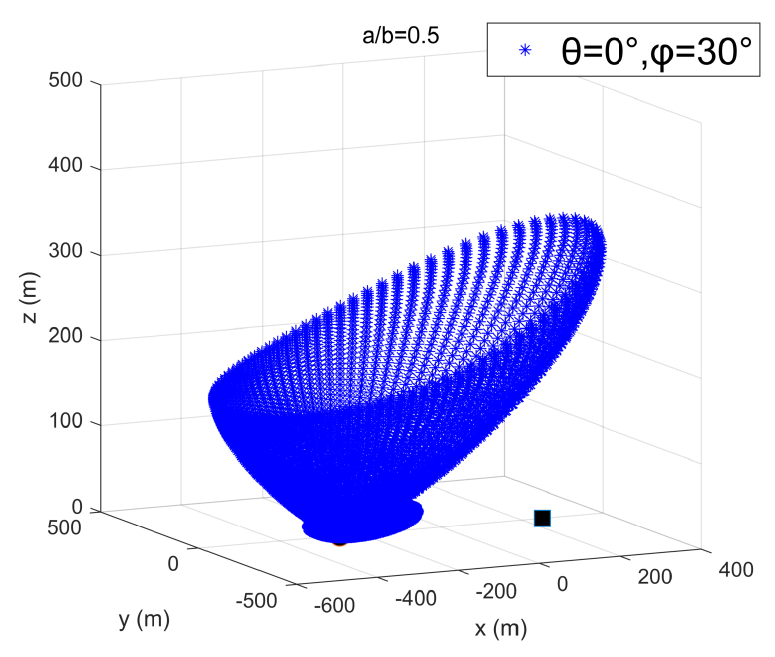}}

	\caption{Oval of Cassini for sensing coverage.}
\end{figure}

\subsection{UAVs for Ground Coverage}

To perform network-level coverage analysis for ground UEs is to cautiously determine the adjustable network parameters according to the metrics of interest so as to optimize the system performance in certain respect. For example, due to the higher probability of NLOS link when operating at a lower altitude as well as the severer potential inter-cell interference when operating at a higher altitude, there might be an optimum deployment altitude for the UAVs so that the coverage probability could be maximized. Similar trade-offs apply to various other parameters like deployment density and number of layers. One favorable mathematical tool for this task is stochastic geometry\cite{banagar2020performance}. By treating the positions of entities in the network as realization of certain random point process, some closed-form functional relationships between network performance metrics and parameters to be optimized, which are rather instructional for practical deployment, can be acquired. In addition, compared to static ground BSs, UAV's 3D mobility creates an additional exploitable degree of freedom for maximizing communication and sensing coverage which is accomplished by optimizing both the trajectory and the beamforming vectors during the platform's flight. Lastly, CKM and other radio map-like concepts\cite{zeng2021simultaneous} can also be of great utility for coverage prediction. All these topics mentioned above can mutually complement one another for common coverage enhancement.

\section{Conclusion}
In this article, we discussed some basic aspects regarding a UAV ISAC network, with an emphasis on UAV's roles as aerial targets to be sensed and aerial anchors facilitating sensing, by highlighting the new characteristics and challenges to be addressed. Furthermore, we pointed out the unique advantages of using cellular network to support UAV ISAC applications and discussed various issues specific to networking. In conclusion, cellular-connected UAV ISAC is conceived to be a promising paradigm for both UAV communication and sky-based sensing, which may help breed new use cases and numerous business opportunities.

\begin{appendices}       

\end{appendices}

\bibliographystyle{IEEEtran}
\bibliography{UAV_ISAC}

\end{document}